\documentclass[12pt]{iopart}
\usepackage{epsfig} 
\usepackage{iopams}
\newcommand{\be}{\begin{eqnarray}}
\newcommand{\ee}{\end{eqnarray}}

\begin{document}

\title{Extended Skyrme interaction (I): spin fluctuations in dense matter}

\author{J. Margueron$^1$ and H. Sagawa$^2$}
\address{$^1$Institut de Physique Nucl\'eaire, IN2P3-CNRS and Universit\'e
Paris-Sud, F-91406 Orsay Cedex, France}
\address{$^2$Center for Mathematical Sciences, University of Aizu,
Aizu-Wakamatsu, 965-8580 Fukushima, Japan}

\begin{abstract}
Most of the Skyrme interactions are known to predict spin or isospin instabilities beyond 
the saturation density of nuclear matter which contradict  
predictions based on realistic interactions.
A modification of the standard Skyrme interaction is proposed so that
the ferromagnetic instability is removed.
The new terms are density dependent and modify only the spin p-h
interaction in the case of spin-saturated system.
Consequences for the nuclear response function and
neutrino mean free path are shown.
The overall effect of the RPA correlations makes dense matter more transparent 
for neutrino propagation by a factor of 2 to 10 depending of the density.
\end{abstract}

%\date{\today}


%\keywords{Skyrme effective interaction, nuclear matter, spin instabilities,
%RPA, neutrino mean free path} 

\submitto{\JPG}

%\maketitle

%%%%%%%%%%%%%%%%%%%%%%%%%%%%%%%%%%%%%%%%%%
%%%%%%%%%%%%%%%%%%%%%%%%%%%%%%%%%%%%%%%%%%

The understanding of dense nuclear matter and its application to nuclear
physics and to astrophysics is strongly related to the properties of in-medium
nuclear interaction.
During the recent years, much efforts went into the fundamental understanding
of the bare nuclear interaction, either through the developement of
effective field theories derived from QCD symmetries~\cite{apel98} 
or through renormalisation group arguments applied to the Lippmann-Schwinger 
equation which generate a phase-equivalent low energy bare interaction named
$v_{\mathrm{low}-k}$~\cite{bog03}.
These theories aim at a separation of the long-range component of the nuclear forces 
dominated by the pion exchange, which is well under control, from the intermediate
and short range components, which are dominated by correlated pion and heavy 
meson exchanges that are poorly known.
A close agreement between the interaction $v_{\mathrm{low}-k}$ and the effective
Gogny interaction~\cite{ber91} have been found concerning the predictions of pairing 
properties in nuclear matter in a wide range of sub-nuclear densities~\cite{sed03}.

Another modelisation of the nuclear interaction in dense matter based on the symmetries
of QCD, the quark-meson coupling model, see Ref.~\cite{gui06} and references therein,
has  given a sound foundation to 
 the Skyrme effective nuclear interaction~\cite{sky56,vau72}.
The modern Skyrme forces are commonly adjusted to the empirical knowledge of 
nuclear matter at saturation density, and sometimes include also theoretical calculations
of asymmetric or pure neutron matter, as well as a set of nuclei described within
Hartree-Fock + BCS or Hartree-Fock Bogoliubov theories.  So far the Skyrme interactions
are  successfully 
 used to describe properties of nuclei such as binding energies, radii and excited states.
Nevertheless, due to its very simple functional form, the standard Skyrme 
interaction is not suited for being extrapolated in astrophysical 
situation such as neutron stars.
Indeed, the mean field solution obtained with such interactions
become unstable for densities larger than the saturation
density $\rho_0$ and also for neutron rich matter.

A recent extensive study of the symmetry energy deduced
from Skyrme interactions~\cite{sto03} has shown that
the symmetry energy becomes negative beyond the saturation 
density of nuclear matter for many of these interactions. 
This is related to the instability of the mean field with respect to 
the isospin density fluctuation $\delta\rho_t$ where $\rho_t=\rho_n-\rho_p$.
Depending on the parameters of the interaction, instabilities can occurs 
in different channels, at various densities and at different isospins.  
It has been shown that there is a limited interval of densities  between 
$\rho_0$ and  3$\rho_0$ for which the mean field is 
stable in symmetric matter and also in neutron matter~\cite{mar02}. 
The stability of the ground state toward small fluctuations 
can be studied within the HF+RPA framework. 
The fluctuations around the mean field are induced 
by the particle-hole interaction,
\be
V_\mathrm{ph}&=&\frac{1}{N_0}\sum_\ell (F_\ell + F_\ell' \tau_1\cdot\tau_2
+G_\ell \sigma_1\cdot\sigma_2 %\nonumber \\
+G_\ell'(\tau_1\cdot\tau_2)(\sigma_1\cdot\sigma_2))
P_l(\cos\theta) \; ,
\ee
in terms of the dimensionless Landau parameters
$F_\ell$, $F_\ell'$, $G_\ell$, $G_\ell'$ and Legendre Polynomials 
$P_l(\cos\theta)$. 
$N_0=g m^*k_F/(2\pi^2\hbar^2)$ is the density of state around the Fermi
energy and $g$ is the degeneracy. 
The matter is stable unless one of these parameters of multipolarity $\ell$
becomes lower than $-2\ell-1$.

\begin{figure}[htb]
\begin{center}
\includegraphics[width=0.9\linewidth]{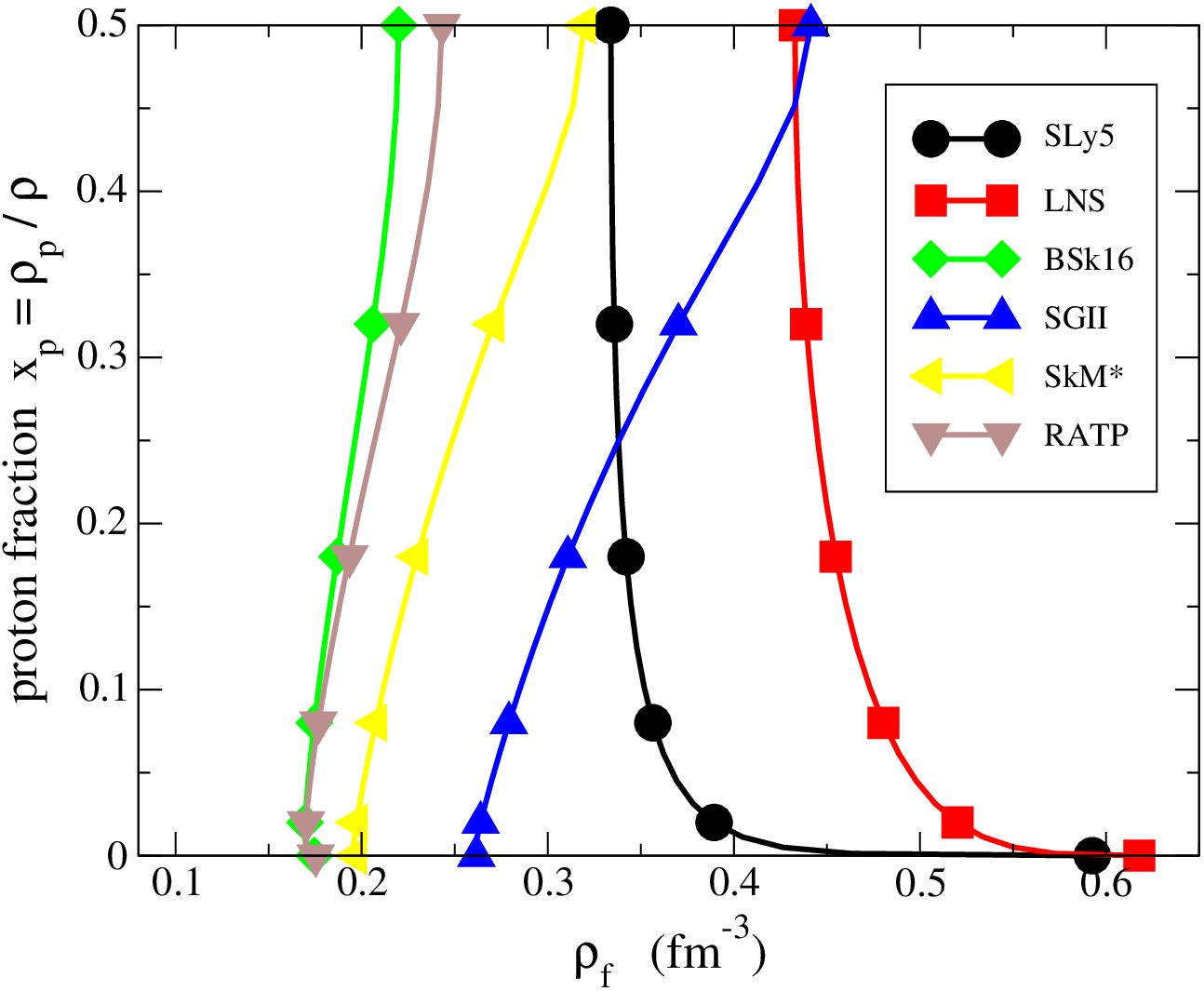}
\end{center}
\caption{Ferromagnetic phase diagram for various effective Skyrme  interactions.
The horizontal axis shows the critical density $\rho_\mathrm{f}$ at which the 
asymmetric matter becomes unstable, while the vertical axis shows the proton 
fraction $\mathrm{x}_p= \rho_p /\rho$.
The matter is spin symmetric for 
smaller density than $\rho_\mathrm{f}$, while it becomes ferromagnetic for larger 
density than $\rho_\mathrm{f}$. 
See the text for details.}
\label{fig1}
\end{figure}

Recently, the ferromagnetic instability has regained interest~\cite{hae96,bha97}
due to the observation of extremely high magnetic fields 
($\sim 10^{15-16}$ G) in compact stars~\cite{tho95} and 
also of giant flares observed recently on 27 December 2004~\cite{laz05}.
We show in Fig.~\ref{fig1} the critical density $\rho_\mathrm{f}$ at which the asymmetric 
matter becomes unstable with respect to the spin fluctuations.  
The proton fraction $\mathrm{x}_p=\rho_p/\rho$ is changed 
from  symmetric matter ($\mathrm{x}_p=1/2$)  to neutron matter ($\mathrm{x}_p=0$). 
We select various  Skyrme interactions which are commonly  used
in the description of finite nuclei. 
Depending on the interaction, the density $\rho_\mathrm{f}$ can be very close
to $\rho_0$ (BSk16~\cite{cha08}, RATP~\cite{RATP}), 
around $2\rho_0$ (SkM$^*$~\cite{SKMS}, SLy5~\cite{Cha97})
or nearly reach $3\rho_0$ (SGII~\cite{SGII}, LNS~\cite{LNS}) in symmetric
nuclear matter ($\mathrm{x}_p$=1/2).
When the proton fraction is decreasing, the density $\rho_\mathrm{f}$ either
decrease (BSk16, RATP, SkM$^*$, SGII) or increase (SLy5, LNS).
From Fig.~\ref{fig1} it could be deduced that
the prediction of $\rho_\mathrm{f}$ is varying largely among  the selected    
Skyrme interactions.

According to microscopic calculations with realistic interactions 
such as diffusion Monte-Carlo~\cite{fan01} or Brueckner Hartree-Fock (HF) 
calculations~\cite{vid02a,vid02b,bom06}, there is no ferromagnetic instability up to 
substantially high densities.
The onset of the ferromagnetic instability beyond a few times $\rho_0$
represents a serious limitation of the Skyrme interaction.
This limitation should be cured in the calculation of nuclear 
matter properties such as equation of state, the response function 
and microscopic processes such as neutrino mean free path.
It is then timely to enrich the effective Skyrme interaction
in order to extend its domain of application and to take 
advantage of its simple form. 
To this end, extended Skyrme interactions shall mimic more accurately 
microscopic results of realistic interactions and preserve the accuracy of the
original Skyrme interaction in the description of nuclei.

It is known that the spin and the spin-isospin excitations have strong impact 
on astronomical observables such as neutrino mean free path in neutron star, 
0$\nu-$ and 2$\nu$ double beta decay processes.  On the other hand, it has 
been recognized also that Skyrme interactions have a serious shortcoming in 
the spin channels. Our main purpose in this manuscript is to improve the spin 
dependent parts of Skyrme interactions  keeping its simplicity and good 
properties for ground state properties. In this way, we will be able to extend 
the field of possible applications of Skyrme interactions for spin dependent 
excitations not only in finite nuclei but also infinite nuclei within the self-consistent 
theoretical model.  It is important to make a bridge between finite and infinite 
systems by the self-consistent model without introducing any extra parameters.  
Top of realistic spin and spin-isospin interactions, we would like to introduce further 
tensor interactions and two-body spin-orbit interactions for constructing a global 
energy density functions of spin and spin-isospin channels.  In this article, we first 
propose a simple way to get ride of the ferromagnetic instability by introducing a  
limited number of new terms to the standard Skyrme interaction in 
Sec.~\ref{sec:I}.
In Sec.~\ref{sec:II}, we apply the new interaction to the calculation
of the RPA response function in nuclear matter 
and then we calculate the neutrino mean free path.
Conclusions are given  in Sec.~\ref{sec:III}.

\section{Extended Skyrme interaction}
\label{sec:I}

\begin{figure}[htb]
\begin{center}
\includegraphics[width=0.95\linewidth]{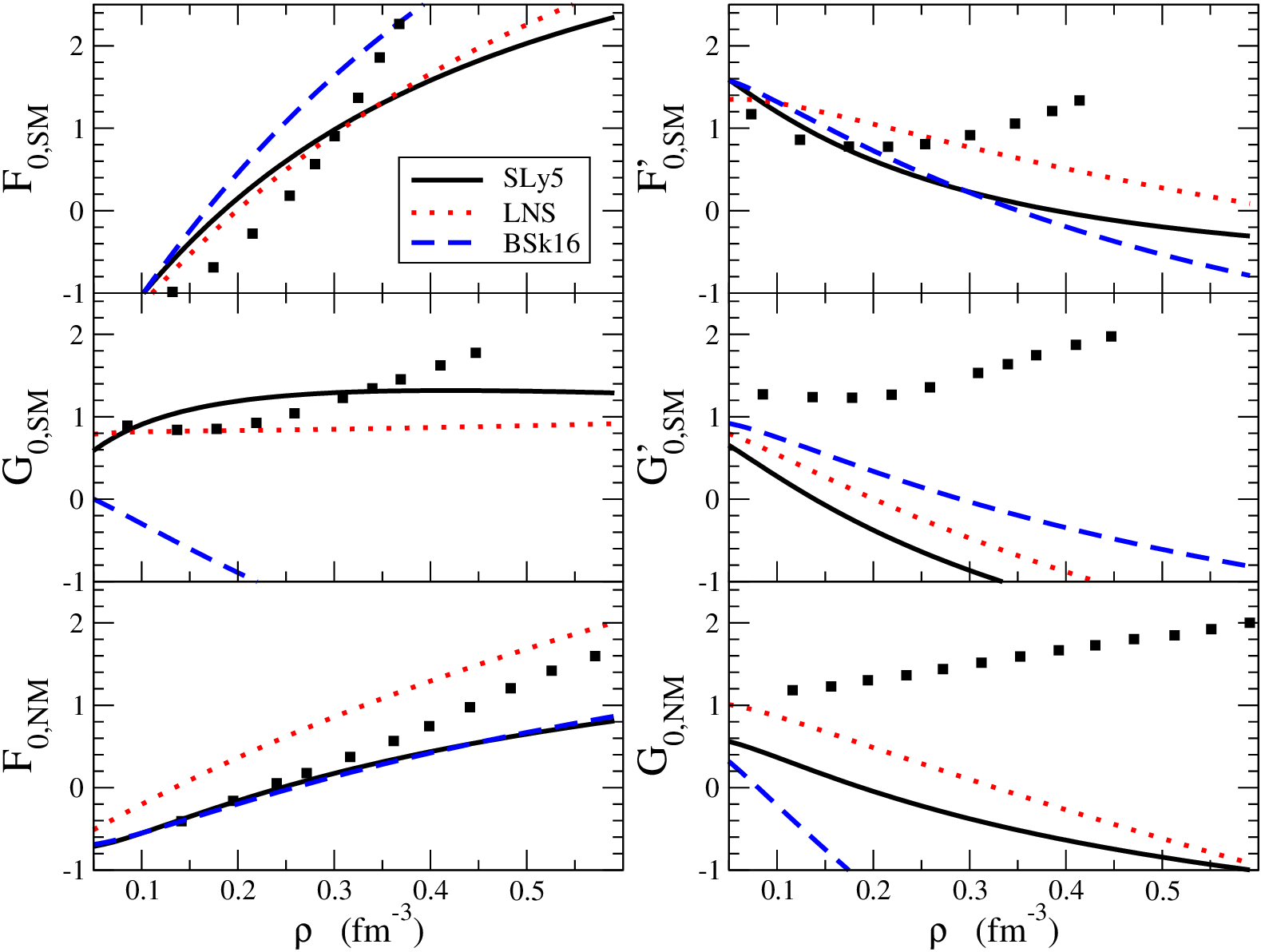}
\end{center}
\caption{Landau parameters in symmetric nuclear matter (SM) and
neutron matter (NM) as a function of the density $\rho$ 
obtained by  the  Skyrme
interactions and the Brueckner HF calculations using 2BF+3BF~\cite{zuo03}. 
Solid line for  SLy5, dotted line for LNS, dashed line for  BSk16 and filled squares for
 Brueckner HF calculations.}
\label{fig2}
\end{figure}

Some of the most recent Skyrme interactions are fitted so as to
reproduce the theoretical binding energy in symmetric and neutron matter
up to a density as much as few times $\rho_0$. 
As a result, the instability in the isospin channel,  which occurs
only if the binding energy of neutron matter becomes more attractive
than that of symmetric matter,  is usually either totally removed or 
pushed to very large densities.
It is then possible to get ride of the isospin instability without
modifying the standard Skyrme interaction.
In the following, we focus on the spin instability and 
test the new terms added to the existing Skyrme 
interactions which do not show  the  instability
in the isospin channel.
We  adopt  three
different  Skyrme interactions, SLy5, LNS and BSk16 for this study. 
SLy5 is an interaction suited by construction to make predictions
 in neutron rich nuclei,
LNS is an interaction which reproduces global features of G-matrix in symmetric
and asymmetric nuclear matter and BSk16 reproduces  known masses of nuclei 
with the best accuracy of  rms deviation  632~keV.
Some nuclear matter 
  properties of these interactions are listed in Tab.~\ref{tab1}.
The monopole Landau parameters in symmetric (SM) and neutron (NM) 
 matter are drawn 
in Fig.~\ref{fig2} as a function of the density.
As expected, these interactions has no instability in the isospin
channel. 
Notice however that in the isospin channel, the Landau parameters $F_\mathrm{0,SM}^\prime$
obtained from the three interaction are more  attractive 
than the one deduced from the microscopic G-matrix calculation~\cite{zuo03} 
(filled squares in Fig.~\ref{fig2}), especially in the high density region.
However, it is not necessary to correct this channel since there are no instabilities
in the density range considered in Fig.~\ref{fig2}.
In the spin channels, the Landau parameters $G_\mathrm{0,SM}$, 
$G_\mathrm{0,SM}^\prime$ and $G_\mathrm{0,NM}$ become more attractive as the density
increases.
The  three Skyrme interactions have  pathological behaviors in the spin
channels  $G_0$ and $G_0^\prime$  
 for which  we now propose a prescription
to cure.

From the density dependence of the Landau parameters represented in Fig.~\ref{fig2},
it is clear that the standard Skyrme interactions are not repulsive enough in the
spin channel, and most probably also in the isospin channel 
at higher density than the saturation density. 
The repulsion shall come from the effect of the three body force~\cite{zuo03}.
Through the density matrix expansion, it has been shown that in-medium many-body 
correlations give rize to density dependent terms in the nuclear 
functional~\cite{neg75}. These terms could be directly included in an effective 
density dependent nuclear interaction such as Skyrme or Gogny.
The most important contribution comes from the scalar-density and it is the only
density which has been considered.
However, spin, isospin and spin-isospin density dependent terms shall also be
considered from the density matrix expansion.
Here, we will  study the impact of the new density dependent terms such as
 the spin and spin-isospin densities:
\be
V^\mathrm{add.}(\mathbf{r}_1,\mathbf{r}_2)=
\frac{1}{6}t_3^s(1+x_3^sP_\sigma)[\rho_s(\mathbf{R})]^{\gamma_s}\delta(\mathbf{r})%\nonumber\\
+\frac{1}{6}t_3^{st}(1+x_3^{st}P_\sigma)[\rho_{st}(\mathbf{R})]^{\gamma_{st}}\delta(\mathbf{r})
\label{eq:addint} \nonumber \\
\ee
where $P_\sigma=(1+\sigma_1\cdot\sigma_2)/2$ is the spin-exchanged operator,
$\mathbf{r}=\mathbf{r}_1-\mathbf{r}_2$ and $\mathbf{R}=(\mathbf{r}_1+\mathbf{r}_2)/2$.
In Eq.~(\ref{eq:addint}), we have introduced the spin density $\rho_s=\rho_\uparrow-\rho_\downarrow$ and
the spin-isospin density
$\rho_{st}=\rho_{n\uparrow}-\rho_{n\downarrow}-\rho_{p\uparrow}+\rho_{p\downarrow}$.
Spin symmetry is satisfied if the power of the density dependent terms $\gamma_s$ and 
$\gamma_{st}$ are even.

In the literature, it has already been proposed to add new terms to the standard  
Skyrme effective interaction.
One of the motivation is that the HF single particle levels are usually too widely 
spaced at the Fermi surface. This shall be corrected by including the coupling of single
particles to vibrations which increase the effective mass of the single particle states
from $m^*/m<1$  to $m^*/m\geq1$ at the Fermi surface.
This dynamical effect could be mimic by introducing a gradient term in the density
dependent interaction (see Refs.~\cite{War83,Liu91} and references therein).
Such a term has also been included in global fits of mass formula 
to make nucleosynthesis calculations as accurate as possible~\cite{Far01}.
However, the removal of the spin instability makes symmetric matter unstable at high 
density~\cite{War79}.
Up to now, there is  no satisfactory additional term which prevent the 
  matter to fall
into spin instabilities.

In the following, we adopt the notations of Ref.~\cite{Cha97} where the density
functional $\mathcal{H}$ is expressed as a sum of the kinetic term $\mathcal{K}$,
a zero range term $\mathcal{H}_0$, a density dependent term $\mathcal{H}_3$,
an effective-mass term $\mathcal{H}_\mathrm{eff}$ and some additional terms
coming from spin-orbit coupling, spin and gradient coupling and coulomb interaction.
The additional terms~(\ref{eq:addint}) modify the density dependent
part,  $\mathcal{H}_3$  to be
$\mathcal{H}_3+\mathcal{H}_3^{s}+\mathcal{H}_3^{st}$,  
where the additional density dependent terms reads
\be
\mathcal{H}_3^s&=&\frac{t_3^s}{12} \rho_s^{\gamma_s}
\Big[ (1+\frac{x_3^s}{2})\rho^2+\frac{x_3^s}{2}\rho_s^2
-(x_3^s+\frac{1}{2})(\rho_n^2+\rho_p^2) %\nonumber \\
-\frac{1}{2}(\rho_{sn}^2+\rho_{sp}^2) \Big] , \\
\mathcal{H}_3^{st}&=&\frac{t_3^{st}}{12}\rho_{st}^{\gamma_{st}}
\Big[ (1+\frac{x_3^{st}}{2})\rho^2+\frac{x_3^{st}}{2}\rho_s^2
-(x_3^{st}+\frac{1}{2})(\rho_n^2+\rho_p^2) %\nonumber \\
-\frac{1}{2}(\rho_{sn}^2+\rho_{sp}^2) \Big] ,
\ee
where $\rho_{sn}=\rho_{n\uparrow}-\rho_{n\downarrow}$ and 
$\rho_{sp}=\rho_{p\uparrow}-\rho_{p\downarrow}$.
The  mean field $U_q$, where $q=n,p$, is then  corrected to be 
\be
U_q^\mathrm{add.}&=&\frac{t_3^s}{12}\rho_s^{\gamma_s}\{(2+x_3^s)\rho-(1+2x_3^s)\rho_q\}
%\nonumber \\
+\frac{t_3^{st}}{12}\rho_{st}^{\gamma_{st}}\{(2+x_3^{st})\rho-(1+2x_3^{st})\rho_q\}.
\label{eq:uq}\nonumber \\
\ee
In symmetric nuclear matter the Landau parameters $G_0$ and $G_0'$~\cite{Nav97} 
are also  modified by the following additional terms
\be
\frac{G_0^\mathrm{add.}}{N_0}  &=& \frac{t_3^s}{48}\gamma_s(\gamma_s-1)
[3\rho^2\rho_s^{\gamma_s-2}-(2x_3^s+1)\rho_t^2-\rho_{st}^2 ] 
+\frac{t_3^{st}}{12}(x_3^{st}-\frac{1}{2})\rho_{st}^{\gamma_{st}}\nonumber \\
&&+\frac{t_3^s}{24}(x_3^s-\frac{1}{2})\left(\gamma_s+1\right)\left(\gamma_s+2\right)\rho_s^{\gamma_s} 
 \; , \label{eq:g0sm}\\
\frac{G_0^{\prime \mathrm{add.}}}{N_0} &=& \frac{t_3^{st}}{48}\gamma_{st}(\gamma_{st}-1)
[3\rho^2+\left(2x_3^s-1\right)\rho_s^2 
-(2x_3^{st}+1)\rho_t^2 ]\rho_{st}^{\gamma_{st}-2}
-\frac{t_3^s}{24}\rho_s^{\gamma_s} \nonumber \\
&&-\frac{t_3^{st}}{48}(\gamma_{st}+2)(\gamma_{st}+1)\rho_{st}^{\gamma_{st}} \; .  
\label{eq:gp0sm}
\ee

In pure neutron matter, the additional terms for $G_0$ are 
\be
\frac{G_0^\mathrm{add.}}{N_0} &=& \frac{t_3^s}{24}(1-x_3^s)\gamma_s(\gamma_s-1)\rho^2\rho_s^{\gamma_s-2} 
+\frac{t_3^{st}}{12}\left(x_3^{st}-1\right)\rho_{st}^{\gamma_{st}} \nonumber \\
&&+\frac{t_3^s}{24}\left(x_3^s-1\right)(\gamma_s+2)(\gamma_s+1)\rho_s^{\gamma_s} \; .
\label{eq:g0nm}
\ee

Non-trivial result for the Landau parameters $G_0^\mathrm{add.}$ and $G_0^{\prime \mathrm{add.}}$
imposes a condition on  
 the powers of the density dependence in Eq.~(\ref{eq:addint})
 to be uniquely fixed as 
$\gamma_s=2$ and $\gamma_{st}=2$.  Notice that spin symmetry is  
also satisfied by these selections.
It turns out that 
 the additional contributions~(\ref{eq:uq}) to the mean field are 
simply null in spin-saturated systems, as the densities $\rho_s=0$ and $\rho_{st}=0$. 
It is thus  possible to add the new
terms~(\ref{eq:addint})  to the  existing Skyrme forces
without destroying good properties of the original Skyrme interactions. 
We obtain  the contributions to the Landau parameters
\be
G_0^\mathrm{add.}  &=& \frac{N_0}{8}t_3^{s}\rho^2  \; , \label{eq:g0sm2}\\
G_0^{\prime \mathrm{add.}} &=& \frac{N_0}{8}t_3^{st}\rho^2 \; , \label{eq:gp0sm2} 
\ee
in spin-saturated
symmetric matter 
and  
\be
G_0^\mathrm{add.} = \frac{N_0}{12}t_3^s(1-x_3^s)\rho^2 \; .
\label{eq:g0nm2}
\ee
 in spin-saturated neutron matter.  
The contributions to the Landau parameters $G_0$ and $G_0'$ in symmetric matter 
in Eq.~(\ref{eq:g0sm2}) and ~(\ref{eq:gp0sm2}) depend only on the 
parameters $t_3^{s}$ and $t_3^{st}$, while that for the Landau parameter 
$G_0$ in neutron matter in Eq.~(\ref{eq:g0nm2})  depends on $t_3^{s}$ and $x_3^{s}$.
We shall then first fix $t_3^s$ and $t_3^{st}$ so as to reproduce the Landau
parameters in symmetric matter, and then we fix $x_3^s$ in neutron matter.
Since the Landau parameters $G_0$ and $G_0^\prime$ are independent of 
$x_3^{st}$, this parameter can be set as $x_3^{st}$=0.

\begin{figure}[htb]
\begin{center}
\includegraphics[width=0.95\linewidth]{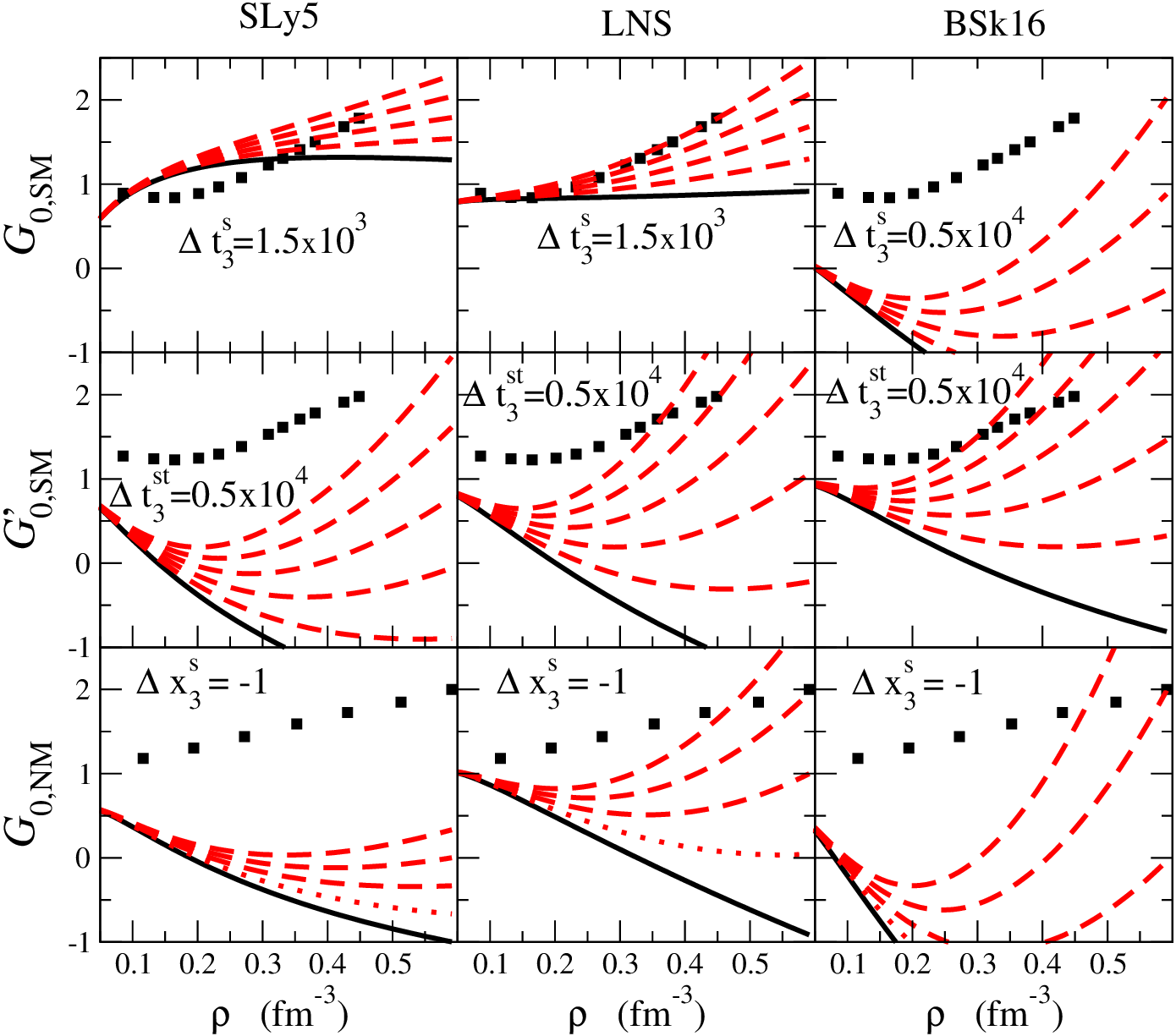}
\end{center}
\caption{Landau parameters for spin and spin-isospin channels 
  $G_0$ and $G_0^\prime$    in symmetric
  nuclear matter and neutron matter. The solid curves are the original  
 ones, while the filled squares are 
  obtained by the  Brueckner HF calculations as  shown in Fig.~\ref{fig2}.
  The dashed curves correspond to different values of the parameters 
  $t_3^s$, $t_3^{st}$ and $x_3^s$.  The parameters are changed from bottom to 
top  multiplying integers (1,2,3,-)  by the mesh 
size given in each window.      
The Landau parameter $G_{0, \mathrm{NM}}$  is calculated with the optimal 
 value for
$t_3^s$  in Tab.~\ref{tab2}, 
   which reproduces best the Brueckner HF $G_{0, \mathrm{SM}}$ values.
The thin dotted lines in the bottom panels correspond  to the results  with 
$x_3^s$=0.}
\label{fig3}
\end{figure}

We show in Fig.~\ref{fig3} the contributions of the new terms for the Landau
parameters $G_0$ and $G_0^\prime$ in symmetric matter and for $G_0$ in neutron matter added to 
 the three effective interactions SLy5, LNS and BSk16.
We compare the original Landau parameters  in Fig.~\ref{fig2} (solid lines)
with the ones which include the contributions of the new terms
(\ref{eq:g0sm}), (\ref{eq:gp0sm}) and (\ref{eq:g0nm}).
The dashed lines correspond to different values of the new parameters 
$t_3^s$, $t_3^{st}$ and $x_3^s$ with the step indicated in each graph.
The values for the parameters $t_3^s$, $t_3^{st}$ and $x_3^s$ resulting from the
best adjustment for asymptotic behavior at very high density
to the Brueckner HF calculations are given in Tab.~\ref{tab2} and are
labelled SLy5st, LNSst and BSk16st respectivelly.
With the parameters given in Tab.~\ref{tab2}, it is confirmed that 
the correction terms to the Landau 
parameters~(\ref{eq:g0sm2})-(\ref{eq:g0nm2}) are repulsive enough 
to remove the spin instabilities.
There are however still differences between the corrected Landau 
parameters and the ones
given by the G-matrix which could be originated from  the tensor interaction.
We can see  that the LNS interaction, fitted originally to the
equation of state deduced from the  G-matrix,  gives the best fitted  Landau 
parameters among the three interactions.
Finally, it could be noticed that the corrected Landau parameters
$G_0^{\prime new}$ are increased by about +0.3 which makes it closer to 
the empirical value.
Notice that the empirical value is estimated to be 1.3$\pm$0.2 from
Wood-Saxon single-particle states plus one-pion and rho meson
models (see for instance Tab.I of Ref.~\cite{ost92} and references therein
and also Ref.~\cite{suz99}) and
0.7$\pm$0.1 from self-consistent Skyrme mean-field models~\cite{frac07}.

\section{RPA response functions and neutrino propagation}
\label{sec:II}

As a consequence of the new terms the spin channels are more
repulsive than  the original Skyrme functional. 
The effect of the new terms could then  modify  the response
function of collective spin modes already at $\rho_0$.
For instance, we show in Tab.~\ref{tab2} the 
Landau parameters $G_0$ and $G_0'$ deduced from the original Skyrme interactions 
and the one including the new terms $G_0^{new}$ and $G_0^{\prime new}$.
The increased of the Landau parameters may have significant influence on 
spin and spin-isospin excitations such as Gamow-Teller states.

\begin{figure}[htb]
\begin{center}
\includegraphics[width=0.95\linewidth]{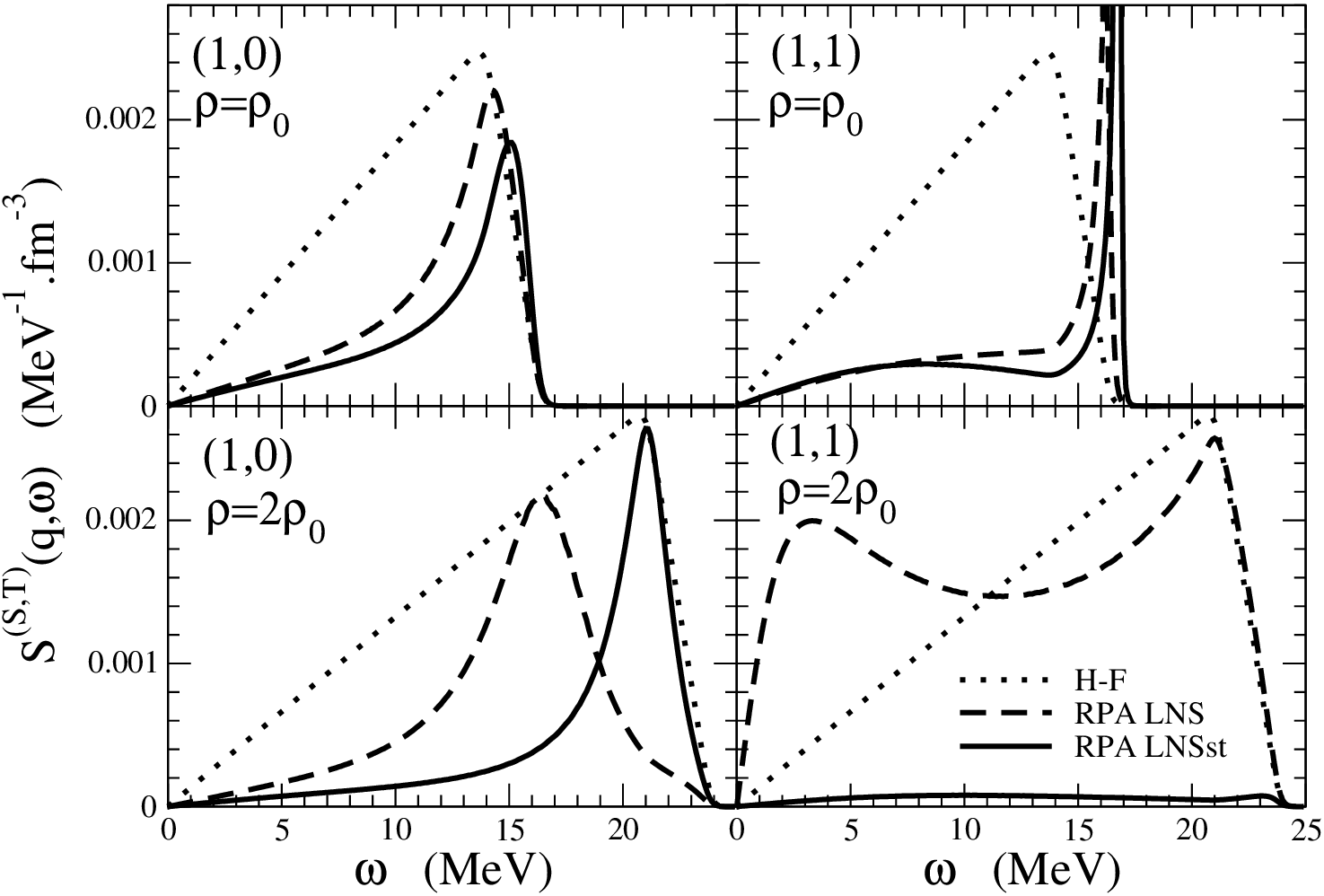}
\end{center}
\caption{RPA response functions $S^{(S,T)}(q,\omega)$ for
  $q$=0.22~fm$^{-1}$ with the temperature $T$=0 MeV calculated with
LNS and LNSst interactions. Top panels are
calculated for $(S,T)=(1,0)$ and $(1,1)$ with $\rho=\rho_0$ and bottom panels
are  for the same channels with $\rho=2\rho_0$.} 
\label{fig4}
\end{figure}

In the following, we calculate the RPA response function in nuclear matter 
at finite temperature,  
\be
S^{(S,T)}(q,\omega,T)=-\frac{1}{\pi}\frac{1}{1-e^{-\omega/T}}{\rm Im}\chi^{(S,T)}(q,\omega,T)
\ee 
where $\chi^{(S,T)}(q,\omega,T)$ is the susceptibility~\cite{fet71}
obtained as the solution of the Bethe-Salpeter equation~\cite{gar92}.
In Fig.~(\ref{fig4}), we show the response functions in the spin
channels ($S=1$) 
at $T=0$ MeV and 
at densities $\rho=\rho_0$ and $\rho=2\rho_0$ calculated by
the LNS and LNSst interactions.
The HF solution (dotted line) is compared with the RPA using the
original LNS Skyrme interaction (dashed line) and also 
with the RPA including the new terms in LNSst (solid line).
According to the semi-classical Steinwedel-Jensen model~\cite{rin82}
the optimal transfered momentum to compare with nuclei shall be
$q=\pi/2R$ where R is the radius of the nuclei. 
For $^{208}$Pb, one thus obtain $q$=0.22~fm$^{-1}$.
For $\rho=\rho_0$, the effect of the new terms is to move the collective mode
to slightly higher energies by 0.5-1 MeV.
For $\rho=2\rho_0$, the $(S=1,T=1)$ Gamow-Teller channel is not far
from being unstable  from the original Skyrme interaction and the low
energy collective mode is being formed.
By the new term $t_3^{st}$,  the low energy mode 
is suppressed and the strength is reduced substantially at $\omega>0$. 
A reduction of the strength at low energy is also observed for the $(S=1,T=0)$ channel.
This effect shall be seen in the calculation of microscopic processes such
as the neutrino cross section.

Neutrinos play a crucial role in physics of supernova 
explosions~\cite{be90} and in the early evolution of their 
compact stellar remnants~\cite{bu86,ja95,red99}. 
To clarify the effect coming from the additional interaction~(\ref{eq:addint}), we consider the case of pure neutron matter.
The scattering of neutrinos on neutrons is then mediated by the neutral current of 
the electroweak interaction. In the non-relativistic limit, the 
mean free path of neutrino with initial energy ${\rm E}_\nu$ is given by~\cite{iwa82}
\begin{eqnarray} 
\lambda^{-1}({\rm E}_\nu,T) &=& \frac{G_F^2}{16 \pi^2} \int d{\bf k}_3  
\Bigg( c_V^2 (1+\cos{\theta})~S^{(S=0)}(q,\omega,T) \nonumber \\
&&+ c_A^2 (3-\cos{\theta})~S^{(S=1)}(q,\omega,T) \Bigg)~,
\label{eq1}
\end{eqnarray}
where $G_F$ is the Fermi constant, $c_V$
($c_A$) is the vector (axial) coupling constant, $k_1=(E_\nu,{\bf k}_1)$ and $k_3$ are
the initial and final neutrino four-momenta, while $q = k_1-k_3$ is the 
transferred four-momentum, and 
$\cos\theta=\hat{\bf k}_1\cdot\hat{\bf k}_3$. 
In the following, we impose the average energy $E_\nu=3T$ in MeV~\cite{red99}.
In Eq.~(\ref{eq1}), the contribution of the response function $S^{(S)}(q,\omega,T)$ 
is clearly identified. 
It describes the response of neutron matter to excitations induced
by neutrinos, and contain the relevant information on the medium.
The vector (axial) part of the neutral current gives
rise to density (spin-density) fluctuations, corresponding to
the  spin $S=0$ ($S=1$) channel.

\begin{figure}[htb]
\begin{center}
\includegraphics[width=0.95\linewidth]{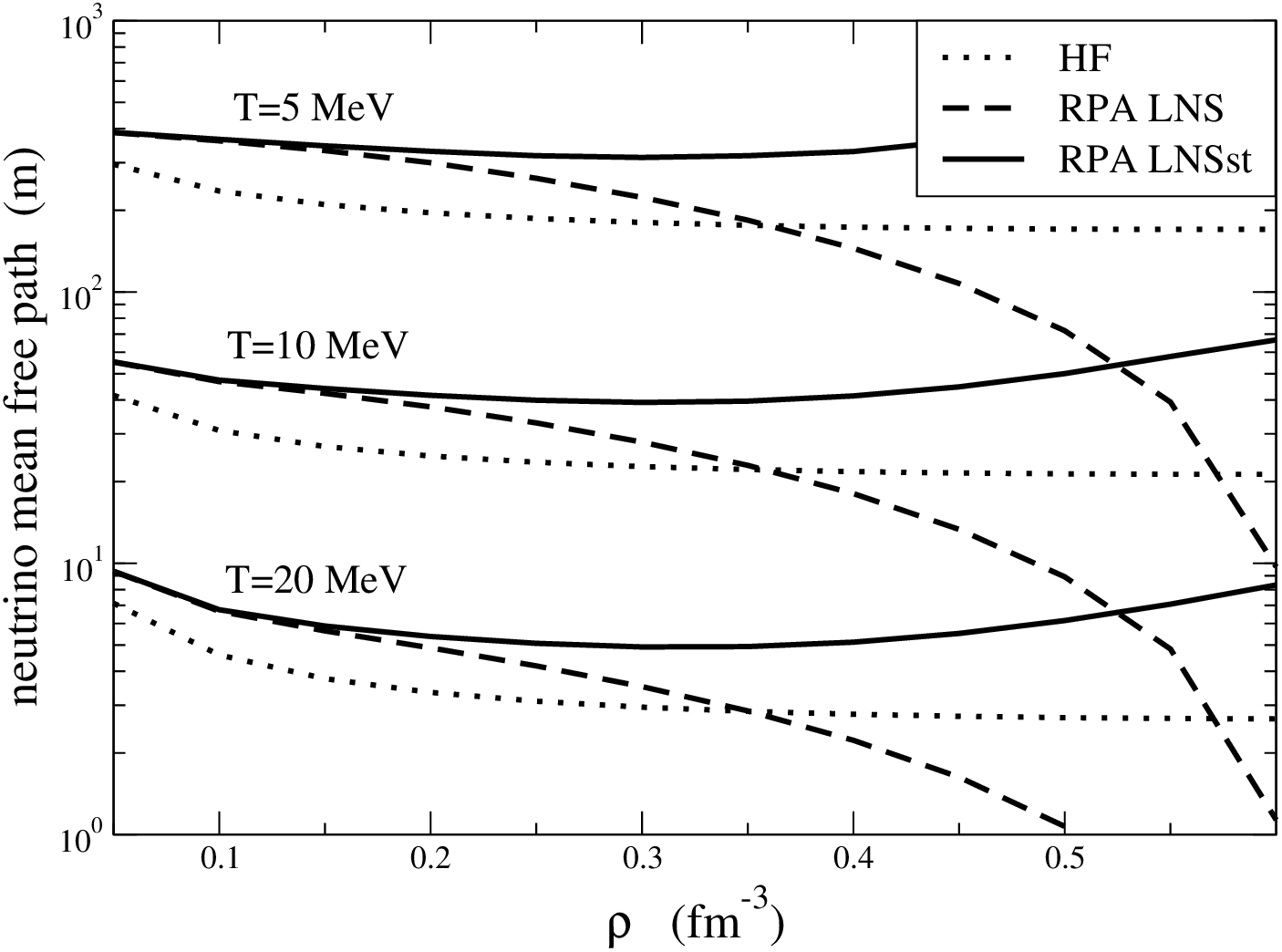}
\end{center}
\caption{Neutrino mean free path in neutron matter for different 
  temperatures $T$=1, 10 and  20 MeV.  The neutrino energy is set to be
  $E_\nu=3T$.
  The mean free path is calculated with the LNS Skyrme
  interaction by the HF mean field approximation (dotted line),
  including the RPA correlations with the 
  original LNS interaction (dashed line) and RPA with the modified spin  
  channel in LNSst (thick line).}
\label{fig5}
\end{figure}

The neutrino mean free path is shown  in Fig.~\ref{fig5} as a function of 
the baryonic density and for different temperatures ($T$=1, 10, 20 MeV).
The neutrino mean free path calculated with the original LNS  Skyrme
interaction (dashed line) is strongly reduced at high density as a
consequence of the onset of the spin instabilities~\cite{her99,mar01}.
The inclusion of the new density dependent interaction~(\ref{eq:addint}) in LNSst
removes the spin instability and reduces the response function as shown in
Fig.~\ref{fig4}. As a consequence, the mean free path is increased
and the matter is more transparent to neutrino as the density increases.
This result is similar to a previous calculation of neutrino mean free
path deduced directly from a microscopic G-matrix~\cite{mar03}.
It illustrates the important contribution of the spin channel to the
neutrino mean free path at high density.
It is also clear from Fig.~\ref{fig5} that the effects of the correlations 
are not washed out by the temperature in the range going up to $T=20$~MeV.

\section{Conclusions}
\label{sec:III}

We have proposed the  extension of the Skyrme 
interaction~(\ref{eq:addint}) which conserves its simplicity 
and extends its domain of stability to large densities and large isospin asymmetries. 
The parameters of the new terms have been adjusted based on microscopic G-matrix calculations.
For the three interactions SLy5st, LNSst and BSk16st, the spin channels become 
   more repulsive compared to the original interactions SLy5, LNS and BSk16; and
 the dimensionless Landau parameter $G_0^{\prime}$ is increased by 
about +0.3.
The simplicity of the extended Skyrme interaction makes possible extensive calculations 
not only in asymmetric and dense matter but also in asymmetric finite nuclei.
We have calculated RPA response functions of spin channels and also
neutrino mean free path in dense matter.
The collapse of the neutrino mean free path at high density is suppressed by the newly
added terms in the Skyrme interaction and
the overall effect of the RPA correlations makes dense matter more transparent 
for neutrino propagation by a factor of  2 to 10 depending on the density.

We would like to extend further the applications of our new spin-density 
dependent interactions for the systematic study of odd mass nuclei, 
the spin-dependent excitations in finite nuclei and also infinite nuclear 
matter. 
Eventually our ultimate goal is to construct the global energy density
functional for the spin and spin-isospin dependent interactions including 
tensor and two-body spin-orbit forces.
Concomitantly, an extensive experimental campaign dedicated to the measurement of
Gamow-Teller  and also other spin-isospin modes is proposed at RIKEN \cite{Sakai2009}. 
The new data will be very usefull to compare with our global energy density
functional.

\ack
%\textbf{Acknowledgments}\\
This work was supported by the Japanese
Ministry of Education, Culture, Sports, Science and Technology
by Grant-in-Aid for Scientific Research under
the Program number C(2) 20540277.

%%%%%%%%%%%%%%%%%%%%%%%%%%%%%%%%%%%%%%%%%%
%%%%%%%%%%%%%%%%%%%%%%%%%%%%%%%%%%%%%%%%%%
%\section*{References}
%\References
%\begin{thebibliography}{00}

\newpage

%%%%%%%%%%%%%%%%%%%%%%%%
\begin{table}[h]
\caption{\label{tab1} Nuclear matter  properties 
   of the Skyrme interactions SLy5, LNS, BSk16
at saturation density $\rho_0$;  the binding energy $a_v$, the
incompressibility $K_\infty$, the symmetry energy $a_s$, 
the density $\rho_\mathrm{f}$ of the ferromagnetic instability.}
\setlength{\tabcolsep}{.1in}
\renewcommand{\arraystretch}{1.2}
\begin{indented}
\item[]
%\begin{center}
\begin{tabular}{cccccc}
%\hline 
\br
& $\rho_0$ & $a_v$ & $K_\infty$ & $a_s$ & $\rho_f/\rho_0$\\
& (fm$^{-3}$) & (MeV) & (MeV) & (MeV ) &  \\
%%$\times\rho_0$ 
%\hline \hline 
\mr
SLy5  & 0.16   & -15.97 & 229.9 & 32   &  2.09\\
LNS   & 0.1746 & -15.32 & 210.9 & 33.4 & 2.48\\
BSk16 & 0.1586 & -16.05 & 241.6 & 30   & 1.24\\
%\hline 
\br
\end{tabular}
%\end{center}
\end{indented}
\end{table}
%%%%%%%%%%%%%

%%%%%%%%%%%%%%%%%%%%%%%%
\begin{table}[h]
\setlength{\tabcolsep}{.05in}
\renewcommand{\arraystretch}{1.3}
\caption{\label{tab2} Parameters of the additional terms $t_3^s$ (in MeV.fm$^{3\gamma_s-2}$),  
$t_3^{st}$ (in MeV.fm$^{3\gamma_{st}-2}$) and  $x_3^s$ 
for the interactions  SLy5st, LNSst and BSk16st  in order to 
reproduce the realistic Landau parameters by Brueckner HF calculations. 
The powers of the density dependence in Eq. (2)  
are set to be $\gamma_s=\gamma_{st}=2$,  and $x_3^{st}=0$.
Are also  shown the dimensionless  Landau parameters $G_0$, $G_0^{new}$, 
$G_0^\prime$ and  $G_0^{\prime \; new}$ 
deduced from the original  Skyrme interactions and  the new interactions 
including the additional terms, respectively.}
%\begin{center}
\begin{indented}
\item[]\begin{tabular}{cccccccc}
%\hline 
\br
& $t_3^s$ & $t_3^{st}$ & $x_3^s$ &  $G_0$ & $G_0^{new}$ & $G_0^\prime$ & $G_0^{\prime \; new}$ \\
%\hline \hline 
\mr
SLy5st & 0.6$\times$10$^4$ & 2$\times$10$^4$ &  -3 & 1.11 & 1.19 & -0.14 & 0.15\\
LNSst  & 0.6$\times$10$^4$ & 1.5$\times$10$^4$ & -1 & 0.83 & 0.95 & 0.14 & 0.45\\
BSk16st & 2$\times$10$^4$ & 1.5$\times$10$^4$ &  -2 &  -0.65 & -0.32 & 0.51 & 0.75\\
%\hline 
\br
\end{tabular}
\end{indented}
%\end{center}
\end{table}
%%%%%%%%%%%%%

\newpage

%%%%%%%%%%%%%
%%%%%%%%%%%%%
%%%%%%%%%%%%%
%%%%%%%%%%%%%
% Obtenu avec critik.f
%%%%%%%%%%%%%
%%%%%%%%%%%%%

%%%%%%%%%%%%%
%%%%%%%%%%%%%
%%%%%%%%%%%%%
% Obtenu avec caminito4 dans cam_sym
%%%%%%%%%%%%%
%%%%%%%%%%%%%

%%%%%%%%%%%%%
%%%%%%%%%%%%%
%%%%%%%%%%%%%
%%%%%%%%%%%%%
% Obtenu avec caminito4 dans cam_sym et caminito1
%%%%%%%%%%%%%
%%%%%%%%%%%%%

%%%%%%%%%%%%%
%%%%%%%%%%%%%
%%%%%%%%%%%%%
%%%%%%%%%%%%%
% Obtenu avec caminito4 dans cam_sym
%%%%%%%%%%%%%
%%%%%%%%%%%%

%%%%%%%%%%%%%
%%%%%%%%%%%%%
%%%%%%%%%%%%%
%%%%%%%%%%%%%
% Obtenu avec caminito1
%%%%%%%%%%%%%
%%%%%%%%%%%%%

\end{document}